\documentclass[twocolumn]{revtex4}

\usepackage{bm}
\usepackage{graphicx,amsmath,amssymb}

\newcommand {\be}  {\begin{equation}}
\newcommand {\ee}  {\end{equation}}
\newcommand {\bea} {\begin{eqnarray}}
\newcommand {\eea} {\end{eqnarray}}

\begin{document}

\title{Gross-Pitaevskii Equation\\ for a System of Randomly Interacting Cold Bosons}
\author{J. van Baardewijk}
\address{Department of Mathematics, King's College University-London, Strand, London,WC2R.2LS,UK.}
\date{\today}

\begin{abstract}
{
Random interaction models have been successful in describing the amorphous properties of solids such as spin-glasses and structural glasses. This modelling approach is applied to a system of zero-spin cold bosons moving in an amorphous environment. The bosons  are given to interact according  to a non-random hard-core interaction.  Additionally the bosons are  subjected  to a random interaction potential similar to that used  for glasses. The approach is to  apply a  combination of replica methods and field theoretic  techniques developed for superfluid Bose systems. This leads to an equation for the low-temperature Bose-Einstein condensate which is derived in the Hartree-Fock approximation. The equation is similar to the Gross-Pitaevskii equation, but the  hard-core coupling constant is renormalised by the presence of the random interactions  in a position dependent way. The amorphous contribution contains the replica diagonal and off-diagonal Green's  functions, for which the Dyson equations are formulated.
}
\end{abstract}

\maketitle

\section{Introduction}
Random interaction potentials are used in models 
of amorphous materials like spin glasses \cite{MezParVir87}, quantum spin glasses  \cite{BrayMoore80,Cugl+01,NieuwRit98} and   quantum glasses \cite{KuHo97,Kuehn03}. Such models are described by a system of interacting distinguishable degrees of freedom. 
The present contribution is concerned with effects such interactions would have for a system of \emph{indistinguishable} degrees of freedom. A system of cold bosons moving in an amorphous background with molecular-scale heterogeneities could be a candidate system described by such a theory if we think of the amorphous background modulating the interactions.

It is interesting to relate and contrast the present project with theories of Bose-glasses \cite{Fisher+89,HuangMeng92,Koba+02} describing superfluid $^4$He confined to nano-porous glasses. The characteristic nano-scale of the disorder in these media (the size of the pores) is considerably larger than the size of the bosons, and is typically described in terms of random external potentials. By contrast, in the present model the amorphous environment is represented by \emph{random-interactions}, and would more realistically require molecular-scale heterogeneity.
As a specific realization one might think of charged bosons moving in an environment with heterogeneous dielectric properties, though it is an open question whether such system would be within reach of current fabrication techniques.
 For the time being our main focus will be on developing the  theoretical description of such systems and to begin elucidating the effects random interactions could have.

We shall formulate the theory in terms of coherent fields and a low-temperature coherent wave-function for which field-theoretic techniques are available, see e.g.   \cite{Bogol47,Popov87,Stoof+09}. In particular this involves constructing a `Gross-Pitaevskii like' equation for the Bose-Einstein condensate \cite{Gross63,Pitaevskii61}. 
The main aim here is to derive this equation and to establish the characteristic manner in which random interactions lead to a renormalisation of the bare (hard-core) coupling between the bosons.

This paper is organised as follows. In section \ref{sec:Ham} the random interaction Hamiltonian is introduced. Section  \ref{sec:Coh}  gives the  partition function in terms of  a path integral in a coherent field representation and defines the disorder average.  Then in section \ref{sec:low-T} the low-temperature condensate wave-function is introduced for which an equation is derived in   
section \ref{sec:Hartree}. Its replica symmetric form is given in section \ref{sec:RSPsi}. Then the Dyson equations for the normal and anomalous Green's functions are given in section \ref{sec:Green}.
Finally, some conclusions are drawn in section \ref{sec:conclusions}.

\section{The model}\label{sec:Ham}
Starting point is the field-Hamiltonian of a three-dimensional system of interacting bosons with zero spin. Such requires the use of creation- and annihilation operators $\hat{\psi}^\dagger(\bm{x})$ and $\hat{\psi}(\bm{x})$, formally creating and annihilating a particle at position $\bm{x}$.  The operators act on number states $|\{n_{\bm{x}}\}\rangle$ in a Fock space with variable  number $N$ of particles. The relations for these operators and the second quantization procedure to construct a Hamiltonian are well known, see e.g. \cite{Feynman72,NegeleOr98}. In particular this assumes an interaction potential function $V(\bm{x},\bm{x'})$ for the interaction of a particle at position $\bm{x}$ with a particle at position $\bm{x'}$, where the vectors $\bm{x}$ and $\bm{x'}$ point to any position in the volume ${\cal V}$. In what follows the three-dimensional integrals $\int d\bm{x}$ are taken over the entire volume.

We choose a random interacting potential of a similar type and interpretation  as the ones used in mean-field glass and spin-glass theory.  
For the field theory proposed here this means that the amorphous background is represented by quenched random couplings $J(\bm{x},\bm{x'})$ describing the interactions of particles at positions $\bm{x}$ and $\bm{x'}$. We will take the $J(\bm{x},\bm{x'})$ to be zero mean Gaussians with
$\overline{J^2(\bm{x},\bm{x'})}=\sigma^2(\bm{x},\bm{x'})$. One would expect $\sigma(\bm{x},\bm{x'})$ to be translationally invariant, $\sigma(\bm{x},\bm{x'})=\sigma(|\bm{x}-\bm{x'}|)$, and decreasing sufficiently rapidly with distance. Alternatively a mean-field model is constructed by choosing $\sigma^2(\bm{x},\bm{x'})=\sigma^2/N$, independently of distance $|\bm{x}-\bm{x'}|$.

 In addition to the random-interaction the bosons are given to interact according a hard-core interaction with (non-random) parameter $V_0$. Such approximation is justified by the fact that at  low temperatures the thermal de-Broglie wavelength of the bosons  is much larger than the range of their interaction.  
In this setting we propose the following (grand canonical) field-Hamiltonian 
\begin{widetext}
\bea\label{eq:fieldHam}
H_J[\hat{\psi}^\dagger,\hat{\psi}]
&=&\,\int d\bm{x}\quad\hat{\psi}^\dagger(\bm{x})\,\Big(-\frac{\nabla_{\bm{x}}^2}{2m}-\mu \Big)\,\hat{\psi}(\bm{x})\nonumber\\
&&+\,\frac{V_0}{2}\int d\bm{x}\int d\bm{x'}\,\,\,\delta(\bm{x}-\bm{x'})\,\hat{\psi}^\dagger(\bm{x})\,\hat{\psi}^\dagger(\bm{x'})\,\hat{\psi}(\bm{x'})\,\hat{\psi}(\bm{x})\nonumber\\
&&+\,\frac{1}{2}\int d\bm{x}\int d\bm{x'}\,\,\, J(\bm{x},\bm{x'})\,\hat{\psi}^\dagger(\bm{x})\,\hat{\psi}^\dagger(\bm{x'})\,\hat{\psi}(\bm{x'})\,\,\,\hat{\psi}(\bm{x}),
\eea
\end{widetext}
where $m$ is the mass of the bosons and $\mu$ the chemical potential. 
The $J$-suffix in $H_J$ emphasises the fact that this Hamiltonian represents a quenched amorphous background characterised by the  $J(\bm{x},\bm{x'})$. 
The factor $N$ is required to assure that the Hamiltonian is  of ${\cal O}(N)$. 
The grand partition function for the configuration $J(\bm{x},\bm{x'})$ is
\be\label{eq:ZJ}
Z_J\,=\,\textrm{Tr}\,\,\textrm{exp}(-\beta H_J[\hat{\psi}^\dagger,\hat{\psi}]\,).
\ee

In order to study the equilibrium properties of the model we need to evaluate the disorder averaged free energy  $F$. The `replica trick' \cite{MezParVir87} allows us to compute it as 
  \be\label{eq:F-bosons}
  -\beta F=\overline{\log{Z}_J}=\lim_{n\to 0}\frac{1}{n}\log{\overline{Z_J^n}},
  \ee
where $n$ represents the number of replicas of the system, for which formally the limit zero is taken.
$Z_J^n$ is the $n$ times replicated partition function. The overline denotes the  integrations over the Gaussians $J(\bm{x},\bm{x'})$, defining the average over all realizations of the random interaction.

\section{Coherent fields}\label{sec:Coh}
To obtain a path integral representation for the partition function in the Matsubara formalism, we use a representation in terms of  coherent states $|\phi\rangle$   and (complex valued) coherent fields $\phi(\bm{x})$ \cite{NegeleOr98}. 
In this representation the replicated partition function for the fixed disorder configuration $J(\bm{x},\bm{x'})$ is 
\be\label{eq:ZR}
Z_J^n\,=\,\int\prod_{a=1}^n {\cal{D}}\phi_a^*{\cal{D}}\phi_a\,\,\,\textrm{exp}{\big(-\frac{1}{\hbar}\,{\cal{S}}_J[\{\phi_a^*,\phi_a\}]\,\big)}.
\ee\\\\
From here on we shall work  in the scaled setting $\hbar=m=1$. 
 The Euclidean action in the coherent field representation  reads
\begin{widetext}
\bea
{\cal{S}}_J[\{\phi_a^*,\phi_a\}]&=&
\sum_a\int_0^\beta d\tau\,\int d\bm{x}\quad\phi_a^*(\bm{x},\tau)\,\Big( \frac{\partial}{\partial \tau}-\frac{1}{2}\nabla_{\bm{x}}^2-\mu \Big)\,\phi_a(\bm{x},\tau)\nonumber\\
&&+\,\frac{V_0}{2}\,\sum_a\int_0^\beta d\tau \int d\bm{x}\quad \phi_a^*(\bm{x},\tau)\,\phi_a^*(\bm{x},\tau)\,\phi_a(\bm{x},\tau)\,\phi_a(\bm{x},\tau)\nonumber\\
&&+\,\frac{1}{2}\sum_a\int_0^\beta d\tau\int d\bm{x}\int d\bm{x'}\,\,\,\,J(\bm{x},\bm{x'})\,\,\,\phi_a^*(\bm{x},\tau)\,\phi_a^*(\bm{x'},\tau)\,\phi_a(\bm{x'},\tau)\,\phi_a(\bm{x},\tau).
\eea
\end{widetext}
Next is to average (\ref{eq:ZR}) over the parameter $J(\bm{x},\bm{x'})$ by means of the following product of  Gaussian integrations 
\bea
\overline{Z_J^n}&=&\int\prod_{(\bm{x},\bm{x'})}\frac{dJ(\bm{x},\bm{x'})}{\sqrt{2\pi\sigma^2(\bm{x},\bm{x'})}}\nonumber\\
&&\textrm{exp}\Big[- \int d\bm{x}\int d\bm{x'}\,\,\frac{J^2(\bm{x},\bm{x'})}{2\sigma^2(\bm{x},\bm{x'})}\Big]\,\,Z_J^n.
\eea
The notation $\prod_{(\bm{x},\bm{x'})}$ means that the integral is over pairs of points, assuming  that $J(\bm{x},\bm{x'})$ is symmetric.
The disorder averaged replicated partition function becomes
\be\label{eq:AvZJn}
\overline{Z_J^n}\,=\,\int\prod_{a=1}^n {\cal{D}}\phi_a^*{\cal{D}}\phi_a\,\,\textrm{exp}{\big(-{\cal{S}}[\{\phi_a^*,\phi_a\}]\,\big)}.
\ee
The final  Euclidean action ${\cal{S}}$ (for the disorder averaged theory) becomes
\begin{widetext}
\bea\label{eq:action}
{\cal{S}}[\{\phi_a^*,\phi_a\}]&=&
\sum_a\int_0^\beta d\tau\int d\bm{x}\quad\phi_a^*(\bm{x},\tau)\,\Big( \frac{\partial}{\partial \tau}-\frac{1}{2}\nabla_{\bm{x}}^2-\mu \Big)\,\phi_a(\bm{x},\tau)\nonumber\\
&&+\,\frac{V_0}{2}\,\sum_a\int_0^\beta d\tau \int d\bm{x}\,\,\,\phi_a^*(\bm{x},\tau)\,\phi_a^*(\bm{x},\tau)\,\phi_a(\bm{x},\tau)\,\phi_a(\bm{x},\tau)\nonumber\\
&&-\,\frac{1}{4}\sum_{ab}\int_0^\beta d\tau \int_0^\beta d\tau'\int d\bm{x} \int d\bm{x'}\,\sigma^2(\bm{x},\bm{x'})\,\phi_a^*(\bm{x},\tau)\,\phi_a^*(\bm{x'},\tau)\,\phi_a(\bm{x'},\tau)\,\phi_a(\bm{x},\tau)\nonumber\\
&&\qquad\qquad\qquad\qquad\qquad\qquad\qquad\qquad\qquad\,\,\,\times\,\,\,\phi_b^*(\bm{x},\tau')\,\phi_b^*(\bm{x'},\tau')\,\phi_b(\bm{x'},\tau')\,\phi_b(\bm{x},\tau').
\eea
\end{widetext}
 In the coherent field representation the particle number constraint  reads
\be\label{eq:canoF}
\int d\bm{x}\,\,\langle\,\phi_a^*(\bm{x},\tau)\phi_a(\bm{x},\tau)  \,\rangle\,=\,N,
\ee
where $\langle ..\rangle$ denotes the average w.r.t. the action ${\cal{S}}$.

\section{Low-temperature condensate}\label{sec:low-T}
Our focus is on the ultra-low temperature behaviour of the system, in particular the effect of Bose-Einstein condensation representing a dominant ground-state occupation.
This is taken into account by performing the shift 
\be\label{eq:shift}
\phi_a(\bm{x},\tau)\,=\,\Psi_a(\bm{x})\,+\,\phi_a'(\bm{x},\tau),
\ee
in the functional integral (\ref{eq:AvZJn})  with 
the assumption that $|\Psi_a(\bm{x})|\gg|\phi_a'(\bm{x},\tau)|$ for typical configurations $\phi_a'(\bm{x},\tau)$ (see e.g. \cite{Bogol47,Popov87,Stoof+09}). The shift $\Psi_a(\bm{x})$ defines  a \emph{non}-fluctuating field and $\phi_a'(\bm{x},\tau)$ is the fluctuating field to be functionally integrated over. The partition function (\ref{eq:AvZJn}) becomes
\be\label{eq:Zshift}
\overline{Z_J^n}\,=\,\int\prod_{a=1}^n {\cal{D}}\phi_a'^*{\cal{D}}\phi_a'\,\,\,\textrm{exp}{\big(-{\cal{S}}^\Psi[\{\phi_a'^*,\phi_a'\}]\,\big)},
\ee
where ${\cal{S}}^\Psi$ denotes the shifted action  
\be\label{eq:SPsi}
{\cal{S}}^\Psi[\{\phi_a'^*\,,\,\phi_a'\}]\,\equiv\,{\cal{S}}\,[\{\Psi_a^*+\phi_a'^*\,,\,\Psi_a+\phi_a'\}].
\ee
The form of ${\cal{S}}^\Psi$ is complicated because $\Psi_a+\phi_a'$ appears to the powers four and eight in the interaction terms of  (\ref{eq:action}). An approximated form  of (\ref{eq:SPsi}) is treated in the next section. 
Averaging (\ref{eq:shift}) gives
\be\label{eq:phiAv} 
\langle\,\phi_a(\bm{x},\tau)\rangle_{S^\Psi}\,=\,\Psi_a(\bm{x})\,+\,\langle\,\phi_a'(\bm{x},\tau)\rangle_{S^\Psi}.
\ee
We require $\Psi_a(\bm{x})=\langle\phi_a(\bm{x},\tau)\rangle_{S^\Psi}$ i.e. $\langle\,\phi'_a(\bm{x},\tau)\,\rangle_{S^\Psi}=0$.
This is achieved by the condition that the \emph{linear} terms  in the shifted action (\ref{eq:SPsi}) vanish.  
Such condition leads to an equation for $\Psi_a(\bm{x})$  similar to the Gross-Pitaevskii equation \cite{Gross63,Pitaevskii61}, derived in the next section.
The particle number constraint (\ref{eq:canoF}) in the shifted theory becomes
\bea\label{eq:cano3}
N\,&=&\,\int d\bm{x}\,\,\,|\Psi_a(\bm{x})|^2+
\int d\bm{x}\,\,\,\langle\phi_a'^*(\bm{x},\tau)\phi_a'(\bm{x},\tau) \rangle\nonumber\\
&\approx&\int d\bm{x}\,\,\,|\Psi_a(\bm{x})|^2.
\eea

\section{Hartree-Fock approximation}\label{sec:Hartree}
Consider the path integral (\ref{eq:Zshift}). The shifted action ${\cal{S}}^\Psi$ may be approximated by  a sum of an action containing only $\Psi$, an action term linear in $\phi'$ and an action term quadratic in $\phi'$
\bea\label{eq:Ssum}
{\cal{S}}^\Psi[\{\phi_a'^*,\phi_a'\}]&=&{\cal{S}}\,[\{\Psi_a^*,\Psi_a\}]\nonumber\\
&&+\,\,{\cal{S}}_{\,\textrm{lin}}^{\,\textrm{(HF)}}[\{\phi_a'^*,\phi_a';\Psi_a^*,\Psi_a\}]\nonumber\\
&&+\,\,{\cal{S}}_{\,\textrm{\footnotesize quadr}}^{\textrm{(HF)}}[\{\phi_a'^*,\phi_a';\Psi_a^*,\Psi_a\}].
\eea
The linear and quadratic term are constructed in the Hartree-Fock  approximation. The quadratic part  will be treated in section \ref{sec:Green} below.
 The linear action ${\cal{S}}_{\,\textrm{\footnotesize lin}}^{\textrm{(HF)}}$ is required  to vanish as mentioned in the previous section. 
It can be written in the general form
\bea\label{eq:linpart}
{\cal{S}}_{\textrm{\footnotesize lin}}^{\,\textrm{\footnotesize (HF)}}&=&
\sum_{a}\int d\tau\int d\bm{x}\,\,\,\Big\{\phi_a'^*(\bm{x},\tau)\,\,{\cal P}_a(\bm{x})\,\Psi_a(\bm{x})\nonumber\\
&&+\,\,\,\phi_a'(\bm{x},\tau)\,\,{\cal P}_a(\bm{x})\,\Psi_a^*(\bm{x})\,\,\Big\}.
\eea
The operator ${\cal P}_a(\bm{x})$  is constructed from the action (\ref{eq:action}) with application of the shift (\ref{eq:shift}). In the Hartree-Fock approximation the result is
\bea\label{eq:P}
{\cal P}_a(\bm{x})&=&\frac{1}{2} \nabla_{\bm{x}}^2-\mu\,+\,V_0\,\Psi_a^*(\bm{x})\Psi_a(\bm{x})\nonumber\\
&&-\sum_{b}\int_0^\beta d\tau'\int d\bm{x'}\,\,G_{ab}(\bm{x'};\tau-\tau')\nonumber\\
&&\times\sigma^2(\bm{x},\bm{x'})\Psi_a(\bm{x'})\Psi_b^*(\bm{x})\Psi_b^*(\bm{x'})\Psi_b(\bm{x}),
 \eea 
where $G_{ab}(\bm{x'};\tau-\tau')$ denotes the Green's function
\be\label{eq:Greens-f}
G_{ab}(\bm{x};\tau,\tau')=\langle \,\phi_a'^*(\bm{x},\tau)\,\phi_b'(\bm{x},\tau')\,\rangle.
\ee
In an equilibrium theory as presented here, the Green's function depends only on the time difference i.e. $G_{ab}(\bm{x};\tau,\tau')=G_{ab}(\bm{x};\tau-\tau')$.
In (\ref{eq:P}) we have exploited the fact that the spatially \emph{off}-diagonal Green's functions $G_{ab}(\bm{x},\bm{x'};\tau-\tau')$   vanish for all $a,b$   \cite{vanBaar09}. 
We have taken only the lowest order Hartree-Fock approximation.
In the second term of (\ref{eq:linpart}) we  integrated by parts twice (w.r.t. $\nabla_{\bm{x}}^2$) using periodic boundary conditions.
Finally, in the amorphous part of (\ref{eq:P})  we omitted a term containing only the functions $\Psi_a(\bm{x}),\Psi_a(\bm{x'}),\Psi_b(\bm{x}),\Psi_b(\bm{x'})$ and no Green's functions. This  is justified by the fact that such term is  ${\cal O}(n)$ which vanishes in the replica limit $n\to 0$.

To make sure that  $\langle\phi_a'(\bm{x},\tau)\rangle=0$ in (\ref{eq:phiAv}) , the coefficients of the linear term (\ref{eq:linpart})  should vanish  i.e.
\begin{subequations}\label{eq:P0}
\begin{align}
 {\cal P}_a(\bm{x})\,\Psi_a(\bm{x})=&0,\label{eq:P0-1}\\
{\cal P}_a(\bm{x})\,\Psi_a^*(\bm{x})=&0.\label{eq:P0-2}
\end{align}
\end{subequations}
The expression ${\cal P}_a(\bm{x})$  defines a real operator since the quartic product of $\Psi$-functions is real valued, so that we only need to consider  (\ref{eq:P0-1}). We further assume that the Green's function $G_{ab}(\bm{x};\tau-\tau')$ is also real valued for all $a,b$. For reference we list here the complete equation (\ref{eq:P0-1}) for $\Psi_a(\bm{x})$
\bea\label{eq:GrossP} 
0&=&\Big[-\frac{1}{2}\nabla_{\bm{x}}^2-\mu\,\,+\,\,V_0\,|\Psi_a(\bm{x})|^2\nonumber\\
&&-\sum_b\int_0^\beta d\tau'\int d\bm{x'}\,\,\,G_{ab}(\bm{x'},\tau-\tau')\nonumber\\
&&\times\,\sigma^2(\bm{x},\bm{x'})\,\Psi_a(\bm{x'})\Psi_b^*(\bm{x})\Psi_b^*(\bm{x'})\Psi_b(\bm{x})\nonumber\\
&&\,\Big]\,\,\Psi_a(\bm{x}).
\eea
This equation defines a `Gross-Pitaevskii like' equation, with the contribution of the term in the second and third line representing the effect of the random interaction potential. In a replica symmetric approximation the structure of this equation will become more transparent, which is subject of section \ref{sec:RSPsi} below.

We also note that in  deriving (\ref{eq:P}), use was made  of the relation 
\be
\int d\bm{ x}\,\,\Psi_a^*(\bm{x})\phi_a'(\bm{x},\tau)+\int d\bm{x}\,\,\Psi_a(\bm{x})\phi_a'^*(\bm{x},\tau)=0.
\ee
 The physical reason behind this condition is  that $\phi_a'(\bm{x},\tau)$ should contain all the configurations that are orthogonal to $\Psi_a(\bm{x})$ \cite{Stoof+09}.

\section{Replica Symmetry}\label{sec:RSPsi}
Next is to write down the  replica symmetric (RS)  equation for the order  parameter $\Psi(\bm{x})$. At this early stage of developing the theory we do not consider replica symmetry breaking.
The RS approximation is defined as 
\begin{subequations}\label{eq:RSdefs}
\begin{align}
\Psi_a(\bm{x})&=\Psi(\bm{x}),\,\,\forall (a)\\
G_{ab}(\bm{x};\tau-\tau')&=G_{d}(\bm{x};\tau-\tau'),\,\,\forall (a=b)\\
G_{ab}(\bm{x};\tau-\tau')&=G(\bm{x}),\,\,\forall (a\ne b)
\end{align}
\end{subequations}
Here we used the Ansatz  that all replica off-diagonal Green's functions are time-independent. The argument to support this is that the replicas are independent and time-translation invariant, and that the origin of time could be chosen independently for each replica. 
In the RS approximation  the equation  (\ref{eq:GrossP}) for $\Psi(\bm{x})$ becomes
\bea\label{eq:RSGrossP} 
0&=&\Big[-\frac{1}{2}\nabla_{\bm{x}}^2-\mu
\,+\,\,V_0\,|\Psi(\bm{x})|^2\nonumber\\
&&-|\Psi(\bm{x})|^2\int_0^\beta d\tau'\int d\bm{x'}\,\,\sigma^2(\bm{x},\bm{x'})\,|\Psi(\bm{x'})|^2\nonumber\\
&&\times\,\big(\,G_d(\bm{x'},\tau-\tau')
-G(\bm{x'})\,\big)\,\,
\,\Big]\,\,\Psi(\bm{x}).
\eea

\subsection{Gross-Pitaevskii equation and effective interactions}
Equation (\ref{eq:RSGrossP}) may be written in the familiar form of the Gross-Pitaevskii equation \cite{Gross63,Pitaevskii61} 
\be\label{eq:RSGrossP-transp} 
0=\Big[-\frac{1}{2}\nabla_{\bm{x}}^2-\mu
\,+\,\,V_{\textrm{eff}}(\beta,\bm{x})\,|\Psi(\bm{x})|^2\,\,\Big]\,\,\Psi(\bm{x}),
\ee
where now the coupling constant $V_{\textrm{eff}}(\beta,\bm{x})$ is an \emph{effective} coupling constant defined as
\bea\label{eq:Veff} 
V_{\textrm{eff}}\,(\beta,\bm{x})&=&V_0
-\int d\bm{x'}\,\,\sigma^2(\bm{x},\bm{x'})\,|\Psi(\bm{x'})|^2\nonumber\\
&&\times\,\beta\,\big(\,\overline{G_d}(\bm{x'})
-G(\bm{x'})\,\big),
\eea
with
\be
\overline{G_d}(\bm{x'})=\frac{1}{\beta}\int_0^\beta d\tau'\,\,G_d(\bm{x'},\tau-\tau'),
\ee
This is one of our main results. Provided only that there is a non-zero condensate wave function $\Psi(\bm{x})\ne 0$ the main effect of random interactions is a renormalisation of the bare (hard-core) interaction in a position dependent way. In the mean-field limit where $\sigma^2(\bm{x},\bm{x'})=\sigma^2/N$, the renormalisation becomes position independent.

Although typical values of the Green's functions are `small' -- they are defined from small fluctuations $\phi'(\bm{x},\tau)$ above the ground-state -- we emphasize the fact that the function $\beta\,(\overline{G_d}(\bm{x'})
-G(\bm{x'}))$ in the disorder contribution of (\ref{eq:Veff}) is  \emph{not} necessarily small, which is  due to the appearance of the factor $\beta$.  In particular at very low temperatures there could be a significant effect from this factor. It is in this low-temperature region where also $|\Psi(\bm{x'})|^2$, appearing in the amorphous term in (\ref{eq:Veff}), is expected to become non-zero.

In section \ref{sec:Green} we shall formulate the Dyson equations for the Green's functions  $G_d$ and $G$. The Green's functions contain the wave-functions $\Psi(\bm{x})$  as we shall see. 
A closed set of equations is obtained by adding to the
Gross-Pitaevskii equation (\ref{eq:RSGrossP-transp}) and the Dyson equations the RS constraint equation (\ref{eq:cano3}) reading
 \be\label{eq:canoRS}
N\,=\,\int d\bm{x}\,\,|\Psi(\bm{x})|^2+
\int d\bm{x}\,\,G_d(\bm{x},0)\approx\int d\bm{x}\,\,|\Psi(\bm{x})|^2,
\ee
defining the normalisation condition for the condensate wave-function $\Psi(\bm{x})$.

\section{Normal and anomalous Green's functions}\label{sec:Green}
The replica symmetric Dyson equations for the normal Green's functions $G_d(\bm{x};\tau-\tau')$ and $G(\bm{x})$ are complicated and involve  anomalous Green's functions.
Anomalous Green's functions are a characteristic feature of theories involving expansions of the kind (\ref{eq:shift}), see e.g. \cite{Popov87,Stoof+09}. We discuss here the general form  of these equations. We shall work again in the Hartree-Fock approximation.
As mentioned above, the normal Green's functions
depend only on the time difference $\tau-\tau'$ and are spatially diagonal.  This applies  also to the anomalous Green's functions since they are determined from the normal Green's functions  by the Dyson equations. 
\\In general the quadratic part of the shifted action (\ref{eq:Ssum}) may be represented   in the terms of a Green's function \emph{matrix} $\bm{{\cal G}}_{ab}(\bm{x};\tau-\tau')$ as
\begin{widetext}
\bea\label{eq:Squadr}
{\cal{S}}_{\,\textrm{\footnotesize quadr}}^{\textrm{(HF)}}[\phi_a'^*,\phi_a';\Psi_a^*,\Psi_a]=\frac{1}{2}\sum_{ab}\int_0^\beta d\tau d\tau'\int d\bm{x} 
\,\begin{pmatrix}
\phi_a'^*(\bm{x},\tau) & \phi_a'(\bm{x},\tau) 
\end{pmatrix}
\,\bm{{\cal{G}}}_{ab}^{-1}(\bm{x};\tau-\tau')
\begin{pmatrix}
\phi_b'(\bm{x},\tau') \\ 
\phi_b'^*(\bm{x},\tau') 
\end{pmatrix}.
\eea
\end{widetext}
The Green's function matrix reads
\bea\label{eq:Gmat}
\bm{{\cal{G}}}_{ab}(\bm{x};\tau-\tau')=
\begin{pmatrix}
G_{ab}(\bm{x};\tau-\tau') & A_{ab}(\bm{x};\tau-\tau') \\\\
\bar{A}_{ab}(\bm{x};\tau-\tau') & G_{ab}(\bm{x};\tau-\tau')
\end{pmatrix},
\eea
where $G_{ab}$ are the \emph{normal} and $A_{ab}$ and $\bar{A}_{ab}$  the \emph{anomalous} Green's functions.  The latter are defined as
\begin{subequations}
\begin{align}
A_{ab}(\bm{x},\tau-\tau')=&\langle \,\phi_a'(\bm{x},\tau)\,\phi_b'(\bm{x},\tau')\,\rangle,\\
\bar{A}_{ab}(\bm{x},\tau-\tau')=&\langle \,\phi_a'^*(\bm{x},\tau)\,\phi_b'^*(\bm{x},\tau')\,\rangle.
\end{align}
\end{subequations}
It should be noted that the normal Green's functions are real valued but  the anomalous Green's functions are \emph{not}.
The Green's functions  satisfy the matrix valued Dyson equations \cite{Popov87,Stoof+09}
\be\label{eq:Dyson}
\bm{{\cal{G}}}_{ab}^{-1}(\bm{x};\tau-\tau')=\bm{{\cal{G}}}_{0,ab}^{-1}(\bm{x};\tau-\tau')
-\,\bm{\Sigma}_{ab}(\bm{x};\tau-\tau'),
\ee
with inverse propagator matrix 
\bea\label{eq:DysonMatrixG0}
&&\bm{{\cal{G}}}_{0,ab}^{-1}(\bm{x};\tau-\tau')=
G_{0,ab}^{-1}(\bm{x};\tau-\tau')
\begin{pmatrix}
1 & 0 \\
0 &  1
\end{pmatrix},
\eea
and inverse propagator
\be\label{eq:freeG}
G_{0,ab}^{-1}(\bm{x};\tau-\tau')=\Big(\frac{\partial}{\partial \tau}-\frac{1}{2}\nabla_{\bm{x}}^2-\mu\Big)\,\delta_{ab}\,\delta(\tau-\tau').
\ee
The self-energy matrix is expressed as
\bea\label{eq:DysonMatrixSigma}
\bm{\Sigma}_{ab}(\bm{x};\tau-\tau')=
\begin{pmatrix}
\Sigma_{ab}^{G}(\bm{x};\tau-\tau')  & \Sigma_{ab}^A(\bm{x};\tau-\tau') \\\\
\Sigma_{ab}^{\bar{A}}(\bm{x};\tau-\tau')  & \Sigma_{ab}^G(\bm{x};\tau-\tau')
\end{pmatrix}.
\eea
Here the superscript $G$ labels the \emph{normal} self-energy and the superscripts $\bar{A}$ and $A$ the \emph{anomalous} self-energies.
They are constructed from (\ref{eq:Squadr}), (\ref{eq:Dyson}), (\ref{eq:DysonMatrixSigma}) and  the action (\ref{eq:action}) with application of the shift (\ref{eq:shift}). In the Hartree-Fock approximation the result is
\bea\label{eq:selfE1}
-\Sigma_{ab}^{G}(\bm{x};\tau-\tau')&=
&\,2V_0\,\,\Psi_a^*(\bm{x})\Psi_a(\bm{x})\,\,\delta_{ab}\,\delta(\tau-\tau')\nonumber\\
&&-2 \int d\bm{x'}\,\,G_{ab}(\bm{x'};\tau-\tau')\sigma^2(\bm{x},\bm{x'}) \nonumber\\
&&\times\,\Psi_a(x')\Psi_a(x)\Psi_b^*(x)\Psi_b^*(x')\nonumber\\
&&- \int d\bm{x'}\,\,G_{ab}(\bm{x};\tau-\tau')\sigma^2(\bm{x},\bm{x'}) \nonumber\\
&&\times\,\Psi_a^*(x')\Psi_a(x')\Psi_b^*(x')\Psi_b(x'),
\eea
\bea\label{eq:selfE2}
-\Sigma_{ab}^A(\bm{x};\tau-\tau')&= 
&\,V_0\,\,\Psi_a(\bm{x})\Psi_a(\bm{x})\,\,\delta_{ab}\,\delta(\tau-\tau')\nonumber\\
&&-\int d\bm{x'}\,\,G_{ab}(\bm{x'};\tau-\tau')\sigma^2(\bm{x},\bm{x'})  \nonumber\\
&&\times\,\Psi_a(x')\Psi_a(x)\Psi_b^*(x')\Psi_b(x),
\eea
\bea\label{eq:selfE3}
-\Sigma_{ab}^{\bar{A}}(\bm{x};\tau-\tau')&=
&\,V_0\,\,\Psi_a^*(\bm{x})\Psi_a^*(\bm{x})\,\,\delta_{ab}\,\delta(\tau-\tau')\nonumber\\
&&-\int d\bm{x'}\,\,G_{ab}(\bm{x'};\tau-\tau')\sigma^2(\bm{x},\bm{x'})\nonumber\\
&&\times\,\Psi_a^*(x)\Psi_a(x')\Psi_b^*(x)\Psi_b^*(x').
\eea
Observe that the normal self-energies are real valued and the anomalous self-energies  complex valued in general. 
The RS representations of the self-energies follow straightforwardly from (\ref{eq:RSdefs}). 
In a first approximation, simplifying the formulation considerably, we may ignore the amorphous parts in the self-energies since the Green's functions  are `small'. 
Finally, in the RS approximation the general form of the Dyson equations (\ref{eq:Dyson}) is \cite{MezParVir87} (in the replica limit  $n\to 0$) 
\begin{subequations}
\begin{align}
(\bm{{\cal{G}}}_d-\bm{{\cal{G}}})^{-1}-\bm{{\cal{G}}}\,(\bm{{\cal{G}}}_d-\bm{{\cal{G}}}
)^{-2}
&=\bm{{\cal{G}}}_{0,d}^{-1}-\bm{\Sigma}_{d},\label{eq:DysonRS1} \\
-\bm{{\cal{G}}}\,(\bm{{\cal{G}}}_d-\bm{{\cal{G}}} )^{-2}&=-\bm{\Sigma}\label{eq:DysonRS2},
\end{align}
\end{subequations}
where (\ref{eq:DysonRS1})  defines the replica diagonal and (\ref{eq:DysonRS2}) the replica off-diagonal part. 
These are matrix equations and constitute the six equations solving for the two normal Green's functions $G_d,G$ and the four anomalous Green's functions $\bar{A}_d,\bar{A}$ and $A_d,A$, each expressed in terms of the wave-function $\Psi(\bm{x})$. 

As mentioned before, the precise  solutions for $\Psi(\bm{x})$ and the Green's functions are currently not yet known. When these become available one may compute the dispersion relations for  collective excitations from the poles of  the replica-\emph{diagonal} Green's function. This amounts to evaluating in momentum space
$
\det{\,\bm{G}_d^{-1}(\bm{k};i\omega_l)}=0
$, 
 where $\omega_l$ are the Matsubara frequencies.
 
\section{Conclusions }\label{sec:conclusions}
A random interaction model for a system of  cold bosons is constructed, applying existing replica-techniques  used  in theories of spin-glasses and structural glasses. The random interaction was assumed to represent an amorphous environment for the bosons. In addition the bosons were given to interact according to a hard-core interaction. A field-Hamiltonian was proposed and  a disorder averaged partition function in the coherent field representation  was presented.
A condition was derived for which a path integral shift was equal to the expectation value of the field. This shift  defined the  wave-function for  the Bose-Einstein  condensate. The condition led to a replica symmetric equation for the wave-function which was similar to the Gross-Pitaevskii equation, the main result of this study so far. 

The hard-core coupling constant was found to be renormalised by the presence of the random interactions in  a temperature and  spatially dependent way, though the spatial dependence of the renormalisation was seen to disappear in a mean-field limit where the variances of the Gaussian couplings are given by $\sigma^2(\bm{x},\bm{x'})=\sigma^2/N$ and loose their spatial characteristics.
The contribution from the random interactions to the equation for the condensate contained the replica diagonal and replica off-diagonal Green's function for which  the  replica symmetric Dyson equations were given. These equations were formulated in terms   normal and anomalous Green's functions. 

Having derived a set of equations to be satisfied by the condensate wave-function and the Green's functions, a next phase in this study is to evaluate these numerically.
Data for the inverse  replica-diagonal Green's function matrix may then   be used to compute the dispersion relation for quasi-particle excitations.
\begin{displaymath}
***
\end{displaymath}

This project has profited considerably from numerous discussions with R. K\"uhn for which I am thankful.


\end{document}